\def\year{2019}\relax
\documentclass[letterpaper, english]{article} 
\usepackage{aaai19}  
\usepackage{times}  
\usepackage{helvet}  
\usepackage{courier}  
\usepackage{url}  
\usepackage{graphicx}  
\usepackage{amsfonts}
\usepackage{amsmath}
\usepackage{amsthm}
\usepackage{amssymb}
\usepackage{babel}
\usepackage[draft]{hyperref}
\hypersetup{colorlinks=false,linkcolor=blue,urlcolor=blue,citecolor=red}
\usepackage{mathtools}
\usepackage{multirow,tabularx}
\usepackage{subfigure} 
\usepackage{balance}

\usepackage{color}

\newcommand{\Lapl}{\mathbf{\mathop{\mathcal{L}}}}
\newcommand{\Trans}[1]{{#1}^{\top}}

\newcommand{\Mat}[1]{\mathbf{#1}}

\newcommand{\Space}[1]{\mathbb{#1}}
\newcommand{\Set}[1]{\mathcal{#1}}

\newcommand{\ie}{\emph{i.e., }}
\newcommand{\eg}{\emph{e.g., }}
\newcommand{\etal}{\emph{et al.}}

\newcommand{\wrt}{\emph{w.r.t. }}

\begin{document}
%
\title{Explainable Reasoning over Knowledge Graphs for Recommendation}

\author{Xiang Wang\textsuperscript{1}\thanks{The first three authors have equal contribution.},
Dingxian Wang\textsuperscript{2}\thanks{Dingxian Wang is the corresponding author.},
Canran Xu\textsuperscript{2},
Xiangnan He\textsuperscript{1,3},
Yixin Cao\textsuperscript{1},
Tat-Seng Chua\textsuperscript{1}\\
\textsuperscript{1}School of Computing, National University of Singapore, \textsuperscript{2}eBay\\
\textsuperscript{3}School of Information Science and Technology, University of Science and Technology of China\\
xiangwang1223@gmail.com,
\{diwang, canxu\}@ebay.com,
\{xiangnanhe, caoyixin2011\}@gmail.com,
dcscts@nus.edu.sg
}

\maketitle
\begin{abstract}
Incorporating knowledge graph into recommender systems has attracted increasing attention in recent years. 
By exploring the interlinks within a knowledge graph, the connectivity between users and items can be discovered as paths,
which provide rich and complementary information to user-item interactions.
Such connectivity not only reveals the semantics of entities and relations, but also helps to comprehend a user's interest.
However, existing efforts have not fully explored this connectivity to infer user preferences, especially in terms of modeling the sequential dependencies within and holistic semantics of a path. 

In this paper, we contribute a new model named \textit{Knowledge-aware Path Recurrent Network} (KPRN) to exploit knowledge graph for recommendation. KPRN can generate path representations by composing the semantics of both entities and relations.
By leveraging the sequential dependencies within a path, we allow effective reasoning on paths to infer the underlying rationale of a user-item interaction.
Furthermore, we design a new weighted pooling operation to discriminate the strengths of different paths in connecting a user with an item, endowing our model with a certain level of explainability.
We conduct extensive experiments on two datasets about movie and music, demonstrating significant improvements over state-of-the-art solutions \textit{Collaborative Knowledge Base Embedding} and \textit{Neural Factorization Machine}.
\end{abstract}

\section{Introduction}
Prior efforts have shown the importance of incorporating auxiliary data into recommender systems, such as user profiles~\cite{TEM} and item attributes~\cite{iCD}.
Recently, knowledge graphs (KGs) have attracted increasing attention~\cite{CKE,RKGE,RippleNet}, due to its comprehensive auxiliary data: background knowledge of items and their relations amongst them.
It usually organizes the facts of items in the form of triplets like (\emph{Ed Sheeran}, \emph{IsSingerOf}, \emph{Shape of You}), which can be seamlessly integrated with user-item interactions~\cite{DBLP:conf/recsys/ChaudhariAM16,DBLP:conf/acl/CaoHJCL17}. 
More important, by exploring the interlinks within a KG, the connectivity between users and items reflects their underlying relationships, which are complementary to user-item interaction data.

\begin{figure}[t]
\includegraphics[width=0.48\textwidth]{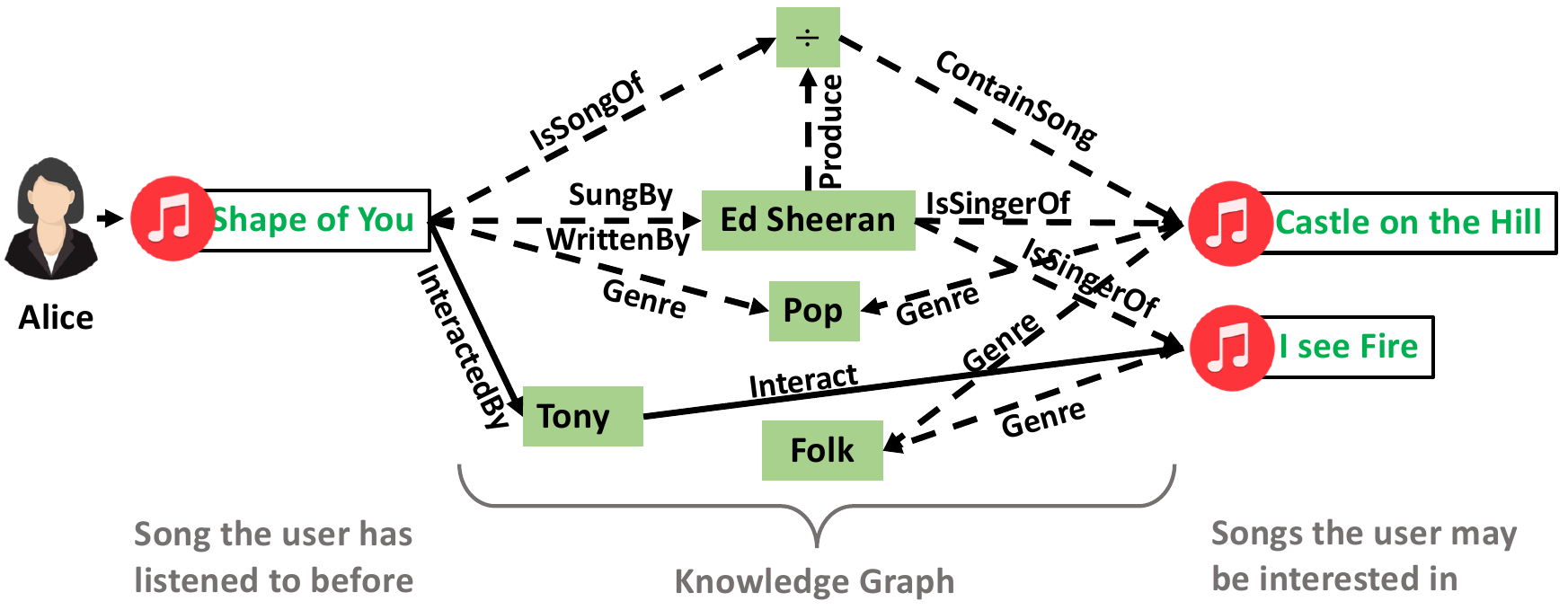}
\vspace{-15pt}
\caption{Illustration of KG-aware recommendation in the music domain. The dashed lines between entities are the corresponding relations, while the sold lines are the user-item interactions.}\label{fig:intro}
\vspace{-15pt}
\end{figure}

Extra user-item \textbf{connectivity} information derived from KG endows recommender systems the ability of \textbf{reasoning} and \textbf{explainability}.
Taking music recommendation as an example (Figure~\ref{fig:intro}), a user is connected to \emph{I See Fire} since she likes \emph{Shape of You} sung by the same singer \emph{Ed Sheeran}.
Such connectivity helps to \textbf{reason} about unseen user-item interactions (\ie a potential recommendation) by synthesizing information from paths.

\noindent\textbf{Running Example:}~(\emph{Alice, Interact, Shape of You})$\wedge$(\emph{Shape of You, SungBy, Ed Sheeran})$\wedge$(\emph{Ed Sheeran, IsSingerOf, I See Fire})$\Rightarrow$(\emph{Alice, Interact, I See Fire}).

\noindent Clearly, the reasoning unveils the possible user intents behind an interaction, offering \textbf{explanations} behind a recommendation.
How to model such connectivity in KGs, hence, is of critical importance to inject knowledge into a recommender systems.

Prior efforts on knowledge-aware recommendation are roughly categorized into path and embedding fashion.
Path-based methods~\cite{DBLP:conf/wsdm/YuRSGSKNH14,yu2013collaborative,DBLP:conf/ijcai/GaoYWZLH18} introduce \emph{meta-paths} to refine the similarities between users and items.
However, we argue that meta-path is inefficient in reasoning over KGs, owing to the following limitations:
1)~As relations are usually excluded from meta-paths, they hardly specify the holistic semantics of paths, especially when similar entities but different relations are involved in a meta-path;
and~2)~They fail to automatically uncover and reason on unseen connectivity patterns, since meta-paths requires domain knowledge to be predefined.

Another line of research~\cite{CKE,DKN,Huang18sigir} leverages knowledge graph embedding (KGE) techniques, such as TransE~\cite{TransE} and TransR~\cite{TransR}, to regularize the representations of items.
As a result, items with similar connected entities have similar representations, which facilitate the collaborative learning of user interests.
Despite performance improvements, we argue that KGE regularization lacks the reasoning ability.
Specially, it only considers direct relations between entities, rather than the multi-hop relation paths as the Running Example shows.
Moreover, the characterization of user-item connectivity is achieved in a rather implicit way, that is, to guide the representation learning, but not to infer a user's preference.

In this work, we aim to fill the research gap by developing a solution that reasons on paths to infer user preferences on items.
In terms of reasoning, we expect our method to model the sequential dependencies of entities and sophisticated relations of a path connecting a user-item pair. 
In terms of explainability, we would like our method to discriminate the different contributions of different paths, when inferring user interests.

Towards this end, we propose a new solution, named Knowledge-aware Path Recurrent Network (KPRN), which not only generates representations for paths by accounting for both entities and relations, but also performs reasoning based on paths to infer user preference.
Specifically, we first extract qualified paths between a user-item pair from the KG, each of which consists of the related entities and relations. We then adopt a Long Short-Term Memory (LSTM) network to model the sequential dependencies of entities and relations. Thereafter, a pooling operation is performed to aggregate the representations of paths to obtain prediction signal for the user-item pair. 
More importantly, the pooling operation is capable of discriminating the contributions of different paths for a prediction, which functions as the attention mechanism~\cite{ACF,PathRNN}.
Owing to such attentive effect, our model can offer path-wise explanations such as \emph{Castle on the Hill is recommended since you have listened to Shape of You sung and written by Ed Sheeran}. We conduct extensive experiments on two datasets to verify our method.

The contributions of this work are threefold:
\begin{itemize}
\item We highlight the importance of performing explicit reasoning on KG to better reveal the reasons behind a recommendation.  
\item We propose an end-to-end neural network model to learn path semantics and integrate them into recommendation.  
\item We contribute a dataset to study KG for recommendation by aligning a MovieLens benchmark with IMDB. We verify our method on the data, and release the data and the codes to facilitate the community working on emerging field of KG-enhanced recommendation.  
\end{itemize}

\section{Knowledge-aware Path Recurrent Network}
\begin{figure*}[t]
\centering
\includegraphics[width=0.97\textwidth]{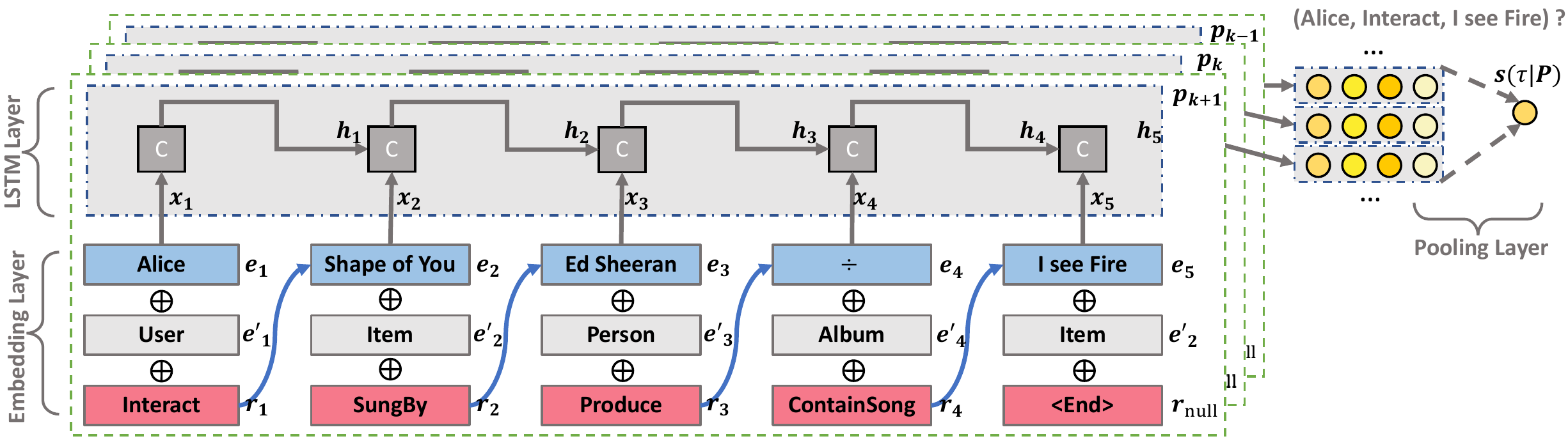}
\vspace{-10pt}
\caption{Schematic overview of our model architecture. The embedding layer contains 3 individual layers for entity, entity type, and relation type, respectively.
The concatenation of the 3 embedding vectors is the input of LSTM for each path.}\label{fig:model}
\vspace{-10pt}
\end{figure*}

In this section, we elaborate our proposed method, as illustrated in Figure~\ref{fig:model}.
Before introducing our proposed method, we first formally define Knowledge Graph, user-item data and describe how to combine them in an enriched knowledge graph as the inputs of our model.


\subsection{Background}
A knowledge Graph (KG) is a directed graph whose nodes are entities $\Set{E}$ and edges $\Set{R}$ denote their relations. Formally, we define KG as $\Set{KG}=\{(h,r,t)| h,r\in\Set{E}, r\in\Set{R}\}$, where each triplet $(h,r,t)$ indicates a fact that there is a relationship $r$ from head entity $h$ to tail entity $t$.

The user-item interaction data is usually presented as a bipartite graph. In particular, we use $\Set{U}=\{u_{t}\}_{t=1}^{M}$ and $\Set{I}=\{i_{t}\}_{t=1}^{N}$ to separately denote the user set and the item set, where $M$ and $N$ are the number of users and items, respectively. Following~\cite{DBLP:conf/recsys/ChaudhariAM16}, we represent the interaction between a user and an item with a triplet $\tau=$($u$, \emph{interact}, $i$), if there is an observed interaction (\eg rate, click, and view feedbacks), where \emph{interact} is a pre-defined relation.

We merge the item set and the entity set through string matching: $\Set{I}\subseteq\Set{E}$, so that the two structural data are integrated into an enriched knowledge graph $\Set{G}=\{(h,r,t)|h,r\in\Set{E}',r\in\Set{R}'\}$, where $\Set{E}'=\Set{E}\cup\Set{U}$ and $\Set{R}'=\Set{R}\cup\{\text{interact}\}$. For consistency, the Knowledge Graph (KG) in the rest paper denotes the combined graph $\Set{G}$ including both original KG and user-item data, otherwise noted.

\subsection{Preference Inference via Paths}
The triplets in the KG clearly describe direct or indirect (i.e. multiple-step) relational properties of items, which shall constitute one or several paths between the given user and item pair. We explore these paths in order to achieve comprehensively reasoning and understanding for recommendation.

Within $\Set{G}$, we formally define the path from the user $u$ to the item $i$ as a sequence of entities and relations: $p=[e_{1}\xrightarrow{r_{1}}e_{2}\xrightarrow{r_{2}}\cdots\xrightarrow{r_{L-1}}e_{L}]$, where $e_{1}=u$, $e_{L}=i$; $(e_{l},r_{l},e_{l+1})$ is the $l$-th triplet in $p$, and $L$ denotes the number of triplets in the path. The construction of paths will be elaborated in the section of~\hyperref[sec:path-extract]{Path Extraction}.

Next, we will use a realistic example to show the sophisticated relations (i.e. paths) between a user and an item behind their possible interactions, which inspires us to model the high-level semantics of path compositionally by considering both entities and (multiple-step) relations.

\vspace{5pt}
\noindent\textbf{Examples:} Consider the music recommendation shown in Figure~\ref{fig:intro}, where the ``listen to Castle on the Hill'' behavior of user Alice can be referred by the following paths:
\begin{itemize}
\item $p_{1}=\text{[Alice}\xrightarrow{\text{Interact}}\text{Shape of You}\xrightarrow{\text{IsSongOf}}\div\xrightarrow{\text{ContainSong}}\text{Castle on the Hill]}$;

\item $p_{2}=~\text{[Alice}\xrightarrow{\text{Interact}}\text{Shape of You}\xrightarrow{\text{SungBy}}\text{Ed Sheeran}\xrightarrow{\text{IsSingerOf}}\text{Castle on the Hill]}$.

\item $p_{3}=~\text{[Alice}\xrightarrow{\text{Interact}}\text{Shape of You}\xrightarrow{\text{InteractedBy}}\text{Tony}\xrightarrow{\text{Interact}}\text{Castle on the Hill]}$;
\end{itemize}
\noindent These paths from the same user \textit{Alice} to the same item \textit{Castle on the Hill} obviously express their different multiple-step relations, and implies various compositional semantics and possible explanations of the listen behavior. In particular, $p_{1}$ and $p_{2} $ infer that \textit{Alice} may prefer songs that belonging to the album $\div$ and the songs sung by \emph{Ed Sheeran}, while $p_{3}$ reflects the collaborative filtering (CF) effect: similar users tend to have similar preferences.
Therefore, from the view of reasoning, we consume the connectivity along all paths to learn compositional relation representations, and weighted pool them together for predicting the \emph{interact} relation between the user and the target item.

\vspace{5pt}
\noindent\textbf{Task Definition:}
Our task can be formulated as follows: given a user $u$, a target item $i$, and a set of paths $\Set{P}(u,i)=\{p_{1},p_{2},\cdots,p_{K}\}$ connecting $u$ and $i$, the holistic goal is to estimate the interaction by:
\begin{align}
\hat{y}_{ui}=f_{\Theta}(u,i|\Set{P}(u,i)),
\end{align}
where $f$ denotes the underlying model with parameters $\Theta$, and $\hat{y}_{ui}$ presents the predicted score for the user-item interaction.
Distinct from embedding-based methods, we can explain $\hat{y}_{ui}$ as the plausibility score of the triplet $\tau=(u,\emph{interact},i)$ inferred by the connectivity $\Set{P}(u,i)$.

\subsection{Modeling}
KPRN takes a set of paths of each user-item pair as input, and outputs a score indicating how possible the user will interact the target item. As illustrated in Figure~\ref{fig:model}, there are three key components: (1) embedding layer to project three types of IDs information: the entity, entity type, and the relation pointing to the next node into a latent space, (2) LSTM layer that encodes the elements sequentially with the goal of capturing the compositional semantics of entities conditioned on relations, and (3) pooling layer to combine multiple paths and output the final score of the given user interacting the target item.

\subsubsection{Embedding Layer}
Given a path $p_{k}$, we project the type (\eg person or movie) and specific value (\eg Peter Jackson or The Hobbit II) of each entity into two separate embedding vectors, $\Mat{e}_{l}\in\Space{R}^{d}$ and $\Mat{e}'_{l}\in\Space{R}^{d}$, where $d$ is the embedding size.

In real-world scenarios, it is common that the same entity-entity pairs may have different semantics due to different relations connecting them.
Such differences may reveal the diverse intents about why a user selected the item.
As an example, let (\emph{Ed Sheeran, IsSingerOf, Shape of You}) and (\emph{Ed Sheeran, IsSongwriterOf, Shape of You}) be the triplets in two paths referring a user's preferences.
Without specifying the relations, these paths will be represented as the same embeddings, regardless of the possibility that the user only prefers songs sung by Ed Sheeran, rather than that written by Ed Sheeran.
We hence believe that it is important to explicitly incorporate the semantics of relations into path representation learning.
Towards this end, each relation $r_{l}$ in $p_{k}$ is represented as an embedding vector $\Mat{r}_{l}\in\Space{R}^{d}$.
As a result, we obtain a set of embeddings for path $p_{k}$, $[\Mat{e}_{1},\Mat{r}_{1},\Mat{e}_{2},\cdots,\Mat{r}_{L-1},\Mat{e}_{L}]$, where each element denotes an entity or a relation. 

\subsubsection{LSTM Layer}
With the embedding sequence to describe a path, we can employ RNN models to explore the sequential information, and generate a single representation for encoding its holistic semantics.
Among various RNN methods, we adopt LSTM since it is capable of memorizing long-term dependency in a sequence.
Such long-term sequential pattern is crucial to reason on paths connecting a user and item entities to estimate the confidence of the ``interact'' relation.

At the path-step $l-1$, the LSTM layer outputs a hidden state vector $\Mat{h}_{l-1}$, consuming the subsequence $[e_{1},r_{1},\cdots,e_{l-1},r_{1-1}]$.
Simultaneously, we concatenate the embedding of current entity $e_{l-1}$ and relation $r_{l-1}$ as the input vector:
\begin{align}\label{equ:triple-rep}
\Mat{x}_{l-1}=\Mat{e}_{l-1}\oplus\Mat{e}'_{l-1}\oplus\Mat{r}_{l-1},
\end{align}
where $\oplus$ is the concatenation operation. Noted that, for the last entity $e_{L}$, a null relation $r_{L}$ is padded to the end of path.
As such, the input vector contains not only the sequential information, but also the semantic information of the entity and its relation to the next entity.
Consequently, $\Mat{h}_{l-1}$ and $\Mat{x}_{l-1}$ are used to learn the hidden state of the next path-step $l$, which is defined via the following equations:
\begin{align}
\Mat{z}_{l}&=\tanh(\Mat{W}_{z}\Mat{x}_{l}+\Mat{W}_{h}\Mat{h}_{l-1}+\Mat{b}_{z})\nonumber\\
\Mat{f}_{l}&=\sigma(\Mat{W}_{f}\Mat{x}_{l}+\Mat{W}_{h}\Mat{h}_{l-1}+\Mat{b}_{f})\nonumber\\
\Mat{i}_{l}&=\sigma(\Mat{W}_{i}\Mat{x}_{l}+\Mat{W}_{h}\Mat{h}_{l-1}+\Mat{b}_{i})\nonumber\\
\Mat{o}_{l}&=\sigma(\Mat{W}_{o}\Mat{x}_{l}+\Mat{W}_{h}\Mat{h}_{l-1}+\Mat{b}_{o})\\
\Mat{c}_{l}&=\Mat{f}_{l}\odot\Mat{c}_{l-1}+\Mat{i}_{l}\odot\Mat{z}_{l}\nonumber\\
\Mat{h}_{l}&=\Mat{o}_{l}\odot\tanh(\Mat{c}_{l})
\nonumber
\end{align}
where $\Mat{c}_{l}\in\Space{R}^{d'}$, $\Mat{z}\in\Space{R}^{d'}$ denote the cell (memory) state vector and information transform module, respectively, and $d'$ is the number of hidden units;
$\Mat{i}_{l}$, $\Mat{o}_{l}$, and $\Mat{f}_{l}$ separately represents the input, output, and forget gate.
$\Mat{W}_{z}$, $\Mat{W}_{i}$, $\Mat{W}_{f}$, and $\Mat{W}_{o}\in\Space{R}^{d'\times 3d}$, and $\Mat{W}_{h}\in\Space{R}^{d'\times d'}$ are mapping coefficient matrices, while $\Mat{b}_{z}$, $\Mat{b}_{i}$, $\Mat{b}_{f}$, and $\Mat{W}_{o}$ are bias vectors.
$\sigma(\cdot)$ is the activation function set as sigmoid, and $\odot$ stands for the element-wise product of two vectors.
Taking advantages of the memory state, the last state $\Mat{h}_{L}$ is capable of representing the whole path $\Mat{p}_{k}$.

Having established the representation of path $\Mat{p}_{k}$, we aim to predict the plausibility of $\tau=(u,\text{interact},i)$.
Towards this end, two fully-connected layers are adopted to project the final state into the predictive score for output, given by:
\begin{align}\label{equ:path-pred}
s(\tau|\Mat{p}_{k})=\Trans{\Mat{W}}_{2}\text{ReLU}(\Trans{\Mat{W}}_{1}\Mat{p}_{k}),
\end{align}
where $\Mat{W}_{1}$ and $\Mat{W}_{2}$ are the coefficient weights of the first and second layers respectively, bias vectors are omitted form simplicity, and the rectifier is adopted as the activation function.

\subsubsection{Weighted Pooling Layer}
A user-item entity pair usually has a set of paths connecting them in a KG.
Let $\Set{S}=\{s_{1},s_{2},\cdots,s_{K}\}$ be the predictive scores for $K$ paths, $\Set{P}(u,i)=\{p_{1},p_{2},\cdots,p_{K}\}$, connecting a user-item pair $(u,i)$, where each element is calculated based on Equation~\eqref{equ:path-pred}.
The final prediction could be the average of the scores of all paths, which is formulated as,
\begin{align}\label{equ:mean-pooling}
\hat{y}_{ui}=\sigma(\frac{1}{K}\sum_{k=1}^{K}s_{k}).
\end{align}

Nevertheless, prior studies~\cite{DBLP:conf/wsdm/YuRSGSKNH14,RKGE} suggest that different paths have varying contributions to model user preferences, while Equation~\eqref{equ:mean-pooling} fails to specify importance of each path.
Inspired by previous work~\cite{reasonchain,ACF}, we design a weighted pooling operation to aggregate scores of all paths.
Here the pooling function is defined as follows,
\begin{align}\label{equ:per-path-score}
g(s_{1},s_{2},\cdots,s_{K})=\log\left[\sum_{k=1}^{K}\exp\left(\frac{s_{k}}{\gamma}\right)\right],
\end{align}
and the final prediction score is given by,
\begin{align}
\hat{y}_{ui}=\sigma(g(s_{1},s_{2},\cdots,s_{K})),
\end{align}
where $\gamma$ is the hyper-parameter to control each exponential weight.
Such pooling is capable of distinguishing the path importance, which is attributed by the gradient:
\begin{align}
\frac{\partial g}{\partial s_{k}}=\frac{\exp(s_{k}/\gamma)}{\gamma \sum_{k'}\exp(s_{k'}/\gamma)},
\end{align}
which is proportional to the score of each path during the back-propagation step.
Moreover, the pooling function endows the final prediction more flexibility.
In particular, when setting $\gamma\rightarrow 0$, the pooling function can degenerate to max-pooling; whereas, it can degrade to mean-pooling by setting $\gamma\rightarrow\infty$.
We conduct a case study on exploring the utility of the weighted pooling operation in Section~\hyperref[sec:case-study]{Case Studies}.

\subsection{Learning}
Similar to the spirit in recent work~\cite{NCF,APR,ItemSilk}, we treat the recommender learning task as a binary classification problem, where an observed user-item interaction is assigned a target value $1$, otherwise $0$.
We use the pointwise learning methods to learn the parameters of our model.
In particular, the negative log-likelihood is adopted as the objective function, which is defined as follows,
\begin{align}
\Lapl=-\sum_{(u,i)\in\Set{O}^{+}}\log\hat{y}_{ui}+\sum_{(u,j)\in\Set{O}^{-}}\log(1-\hat{y}_{uj}),
\end{align}
where $\Set{O}^{+}=\{(u,i)|y_{ui}=1\}$ and $\Set{O}^{-}=\{(u,j)|y_{uj}=0\}$ are the positive and negative user-item interaction pairs, respectively.
We conduct $L_{2}$ regularization on the trainable parameters $\Theta$, which is omitted here for simplicity, to avoid overfitting.
We elaborate the implementation details in the section of~\hyperref[sec:experimental-set]{Experimental Settings}.	
\section{Experiments}
In this section, we perform experiments on two real-world datasets to evaluate our proposed method.
We aim to answer the following research questions:
\begin{itemize}
\item \textbf{RQ1:}~Compared with the state-of-the-art KG-enhanced methods, how does our method perform?
\item \textbf{RQ2:}~How does the multi-step path modeling (\eg the incorporation of both entity and relation types) affect KPRN?
\item \textbf{RQ3:}~Can our proposed method reason on paths to infer user preferences towards items?
\end{itemize}

\subsection{Dataset Description}
We consider two scenarios: movie recommendation and music recommendation.
For movie domain, we use the combination of MovieLens-1M\footnote{\tiny{\url{https://grouplens.org/datasets/movielens/1m/}.}} and IMDb\footnote{\tiny{\url{https://www.imdb.com/}.}} datasets, named MI, which are linked by the titles and release dates of movies.
In particular, MovieLens-1M offers the user-item interaction data, while IMDb serves as the KG part that contains auxiliary information on movies, such as genre, actor, director, and writer.
For music domain, we use the benchmark dataset, KKBox, which is adopted from the WSDM cup 2018 Challenge\footnote{\tiny{\url{https://wsdm-cup-2018.kkbox.events/}.}} and is provided by the music streaming service KKBox.
Beyond the user-item interaction data, this dataset contains the descriptions of music like singer, songwriter, and genre.
The statistics of two datasets are summarized in Table~\ref{tab:data-stat}.

\begin{table}[t]
\centering
\caption{Statistics of our datasets.}
\label{tab:data-stat}
\resizebox{0.47\textwidth}{!}{\begin{tabular}{c|l|r|r}
\hline
\multicolumn{1}{l|}{} & \multicolumn{1}{c|}{Dataset} & \multicolumn{1}{c|}{MI} & \multicolumn{1}{c}{KKBox} \\ \hline\hline
\multirow{3}{*}{\begin{tabular}[c]{@{}c@{}}User-Item\\ Interaction\end{tabular}} & \#Users & 6,040 & 34,403 \\
 & \#Items & 3,859 & 2,296,833 \\
 & \#Interactions & 998,034 & 3,714,655 \\ \hline\hline
\multirow{4}{*}{\begin{tabular}[c]{@{}c@{}}Knowledge\\ Graph\end{tabular}} & \#Entities & 11,462 & 2,851,220 \\
 & \#Entity Types & 4 & 4 \\
 & \#Relation Types & 6 & 6 \\
 & \#Triplets & 1,017,030 & 11,182,682 \\ \hline\hline
\multirow{2}{*}{Path} & \#Paths & 55,573,556 & 38,192,484 \\
 & Avg Path Length & 5.07 & 5.09 \\ \hline
\end{tabular}}\vspace{-15pt}
\end{table}

Following previous efforts~\cite{DBLP:conf/wsdm/YuRSGSKNH14,NCF,RKGE}, we process the datasets as: if a user rates a movie or has an interaction record with a song, we set the user-movie or user-song pair as the observed positive feedback with the target value of $1$, and $0$ otherwise.

For each dataset, we holdout the $80\%$ and $20\%$ interaction history of each user randomly to construct the training and test sets.
For each positive user-item interaction pair in the training set, we conducted the negative sampling strategy to pair it with four negative items that the user has not interacted with.
During the test stage, the ratio between positive and negative interactions is set as $1:100$, namely, $100$ negative items are randomly sampled and pair with one positive item in the testing set.

\subsection{Path Extraction}\label{sec:path-extract}
In practice, it is labor intensive and infeasible to fully exploring all connected paths over the KG.
Especially, the number of paths grows exponentially \wrt the length of path, where millions of interlinks will be generated.
As suggested in prior efforts~\cite{Pathsim,RKGE}, truncating all paths at a certain length and disregarding remote connections are sufficient to model the connectivity between a user-item pair.
Moreover, as pointed out by~\cite{Pathsim}, paths with length greater than six will introduce noisy entities.
Therefore, we extract all qualified paths, each with length up to six, that connect all user-item pairs.

\subsection{Experimental Settings}\label{sec:experimental-set}
\subsubsection{Evaluation Metrics}
We adopt two evaluation protocols to evaluate the performance of top-$K$ recommendation and preference ranking, respectively, given by:
\begin{itemize}
\item \textbf{hit}@$K$ considers whether the relevant items are retrieved within the top $K$ positions of the recommendation list.
\item \textbf{ndcg}@$K$ measures the relative orders among positive and negative items within the top $K$ of the ranking list.
\end{itemize}
We report the average metrics at $K=\{1,2,\cdots,15\}$ of all instances in the test set.

\subsubsection{Baselines}
We compare our proposed method with the following methods:
\begin{itemize}
\item \textbf{MF}~\cite{BPR}: This is matrix factorization with Bayesian personalized ranking (BPR) loss, which solely utilizes user-item interaction.
\item \textbf{NFM}~\cite{NFM}: The method is a state-of-the-art factorization model which treats historical items as the features of users. Specially, we employed one hidden layer as suggested in~\cite{NFM}.
\item \textbf{CKE}~\cite{CKE}: Such embedding-based method is tailored for KG-enhanced recommendation, which integrates the representations from Matrix Factorization~\cite{BPR} and TransR~\cite{TransR} to enhance the recommendation.
\item \textbf{FMG}~\cite{FMG}: This is a state-of-the-art meta-path based method, which predefines various types of meta-graphs and employs Matrix Factorization on each meta-graph similarity matrix to make recommendation.
\end{itemize}

\subsubsection{Parameter Settings}
For fair comparison, we learn all models from scratch without any pretrained parameters.
We optimize all models with Adaptive Moment Estimation (Adam) and apply a grid search to find out the best settings of hyperparameters.
The learning rate is searched in $\{0.001, 0.002, 0.01, 0.02\}$, while the coefficient of $L_{2}$ regularization is tuned amongst $\{10^{-5},10^{-4},10^{-3},10^{-2}\}$.
Other hypermeters of our proposed model are empirically set as follows: the batch size is $256$, the embedding size of relation and entity type is $32$, the embedding size of entity value is $64$, and the unit number of LSTM is $256$.
The dimensions of latent factors for MF, NFM, and CKE are empirically set to be $64$.
For FMG, we set the rank used to factorize meta-graph similarity matrices to be $10$, and the factor size of the second-order weights as $10$, as suggested by~\cite{FMG}.
Moreover, the early stopping strategy is performed, \ie premature stopping if hit@$15$ on the test data does not increase for five successive epochs.

\subsection{Performance Comparison (RQ1)}

\begin{figure*}[t]
\centering
\subfigure[hit@$K$ on MI]{
\label{fig:all-hit-mi}\includegraphics[width=0.24\textwidth]{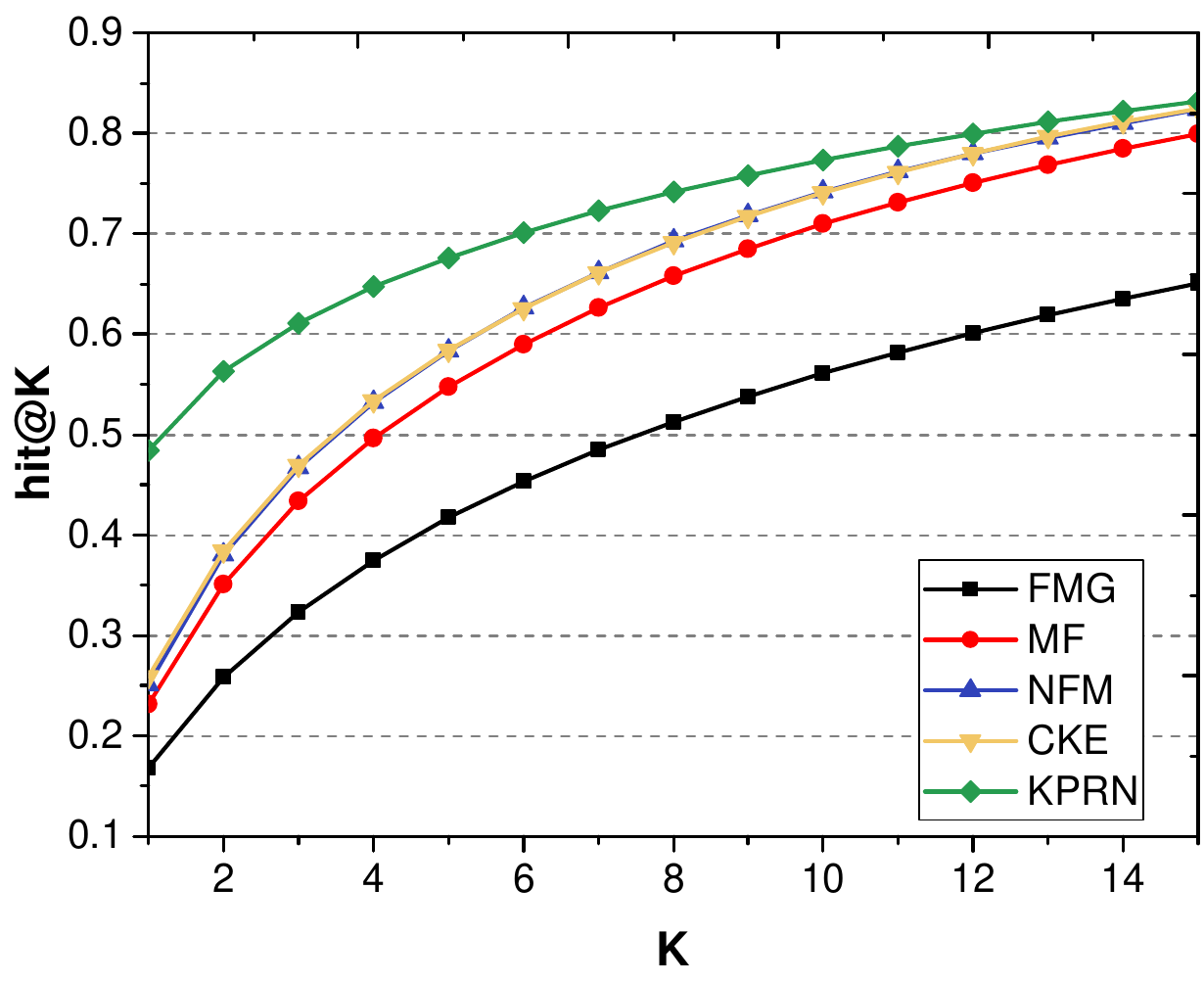}}
\subfigure[ndcg@$K$ on MI]{
\label{fig:all-ndcg-mi}\includegraphics[width=0.24\textwidth]{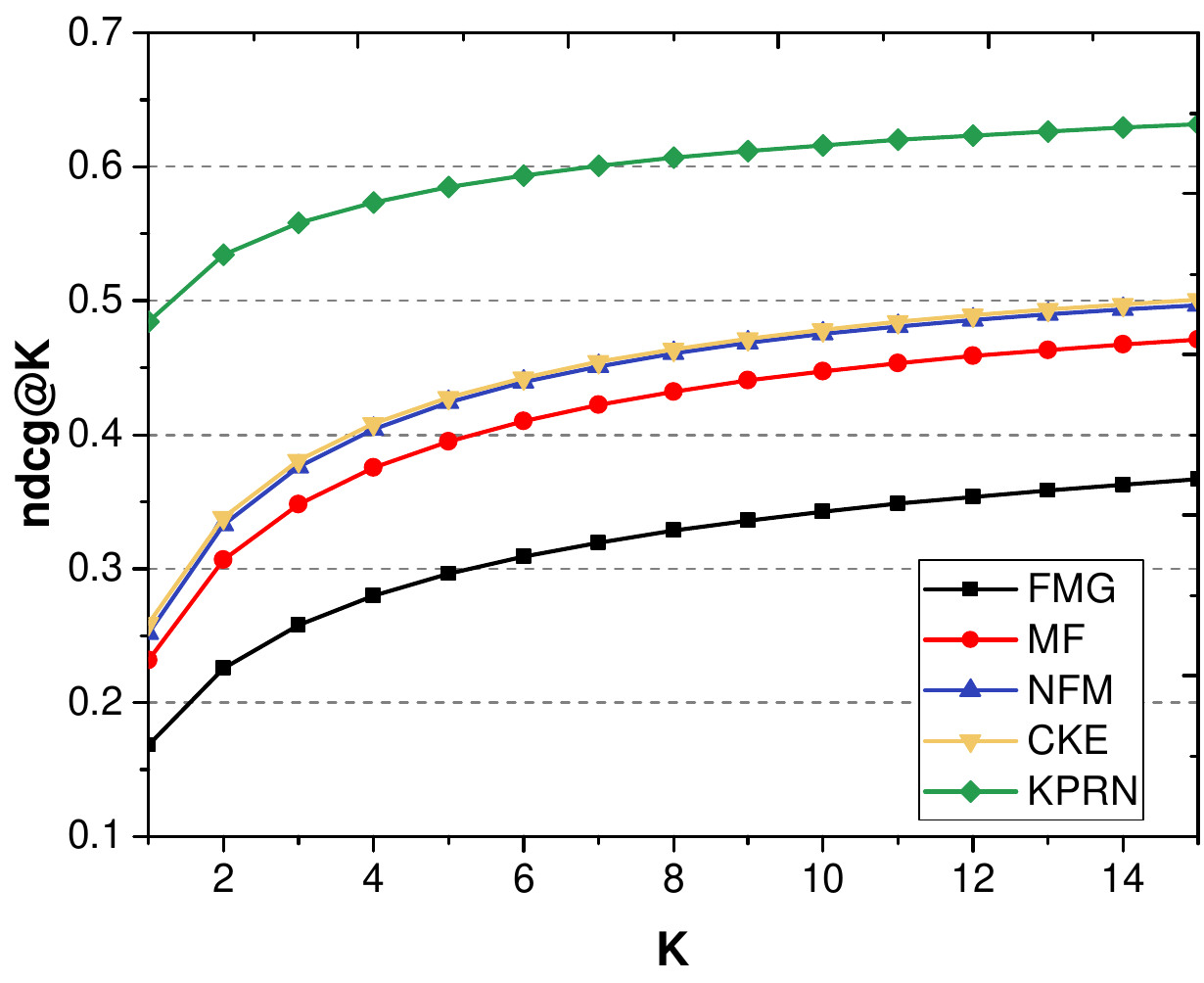}}
\subfigure[hit@$K$ on KKBox]{
\label{fig:all-hit-kkbox}\includegraphics[width=0.24\textwidth]{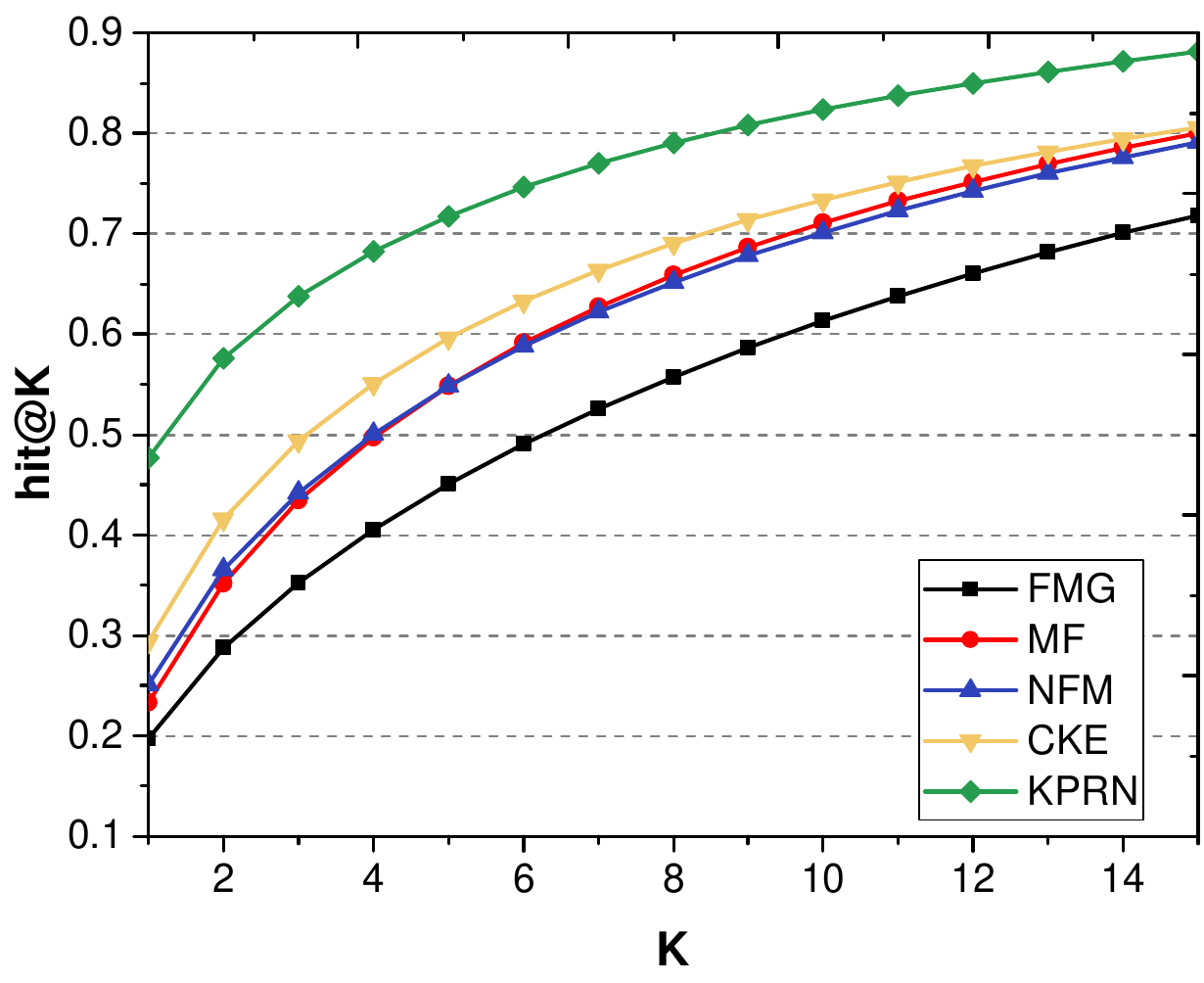}}
\subfigure[ndcg@$K$ on KKBox]{
\label{fig:all-ndcg-kkbox}\includegraphics[width=0.24\textwidth]{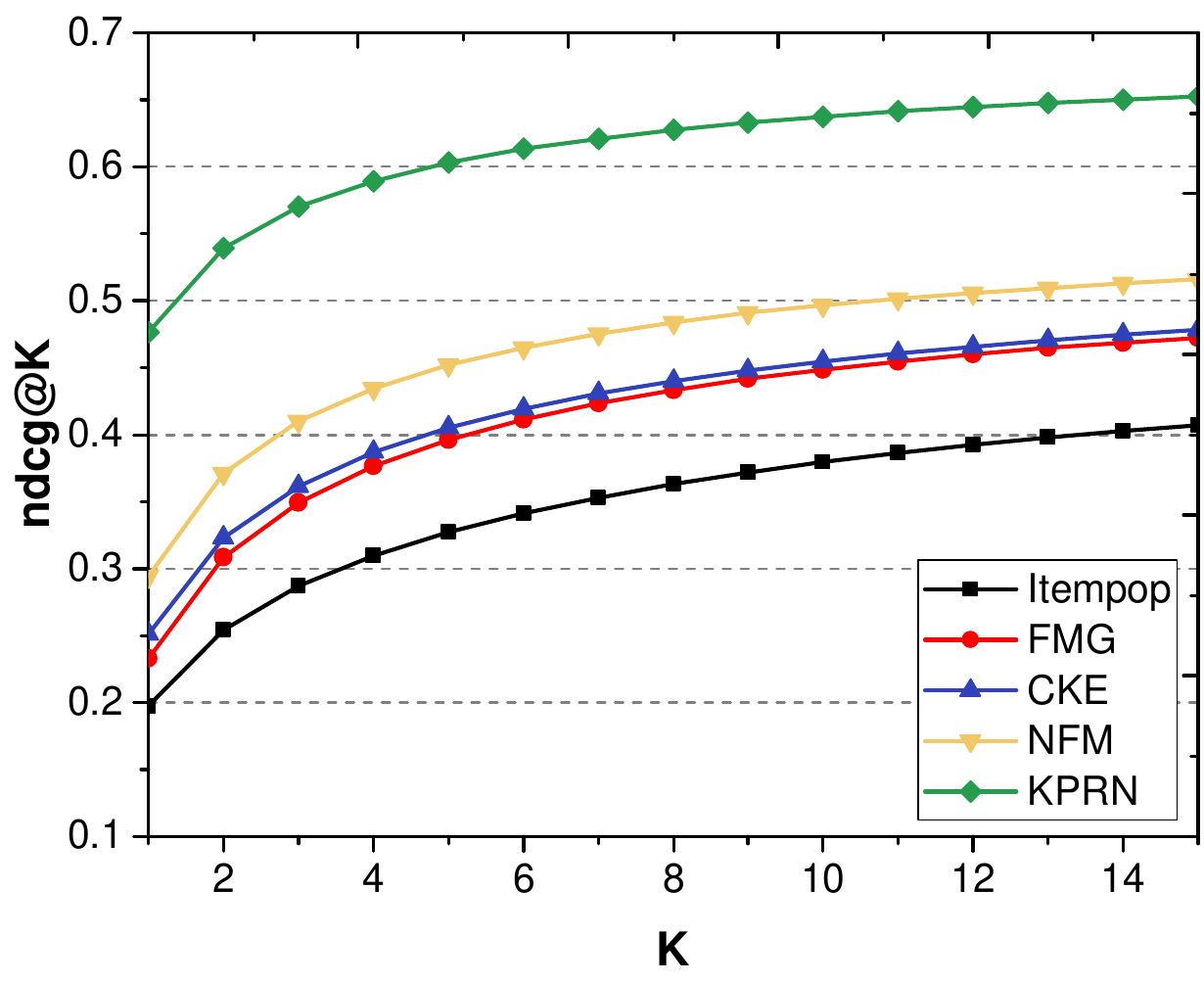}}
\vspace{-15pt}
\caption{Top-$K$ recommendation performance between all the methods on MI and KKBox datasets \wrt hit@$K$ and ndcg@$K$, where $K=\{1,2,\cdots,15\}$.}
\label{fig:overall-comparison}\vspace{-15pt}
\end{figure*}

\begin{table*}[t]
\centering
\caption{Performance comparison of KPRN and KPRN-r and their effects on relation modeling.}
\label{tab:relation-modeling}
\resizebox{0.97\textwidth}{!}
{\begin{tabular}{l|ccc|ccc||ccc|ccc}
\hline
 & \multicolumn{6}{c||}{MI} & \multicolumn{6}{c}{KKBox} \\ \hline
 & hit@5 & hit@10 & hit@15 & ndcg@5 & ndcg@10 & ndcg@15 & hit@5 & hit@10 & hit@15 & ndcg@5 & ndcg@10 & ndcg@15 \\ \hline\hline
KPRN-r & 0.635 & 0.738 & 0.801 & 0.533 & 0.566 & 0.583 & 0.712 & 0.821 & 0.878 & 0.607 & 0.632 & 0.647 \\
KPRN & 0.676 & 0.773 & 0.832 & 0.584 & 0.616 & 0.632 & 0.717 & 0.823 & 0.881 & 0.613 & 0.637 & 0.652 \\ \hline
\end{tabular}}
\vspace{-15pt}
\end{table*}

\begin{figure}[ht]
\centering
\subfigure[hit@$K$ on MI]{
\label{fig:gamma-hit-mi}\includegraphics[width=0.22\textwidth]{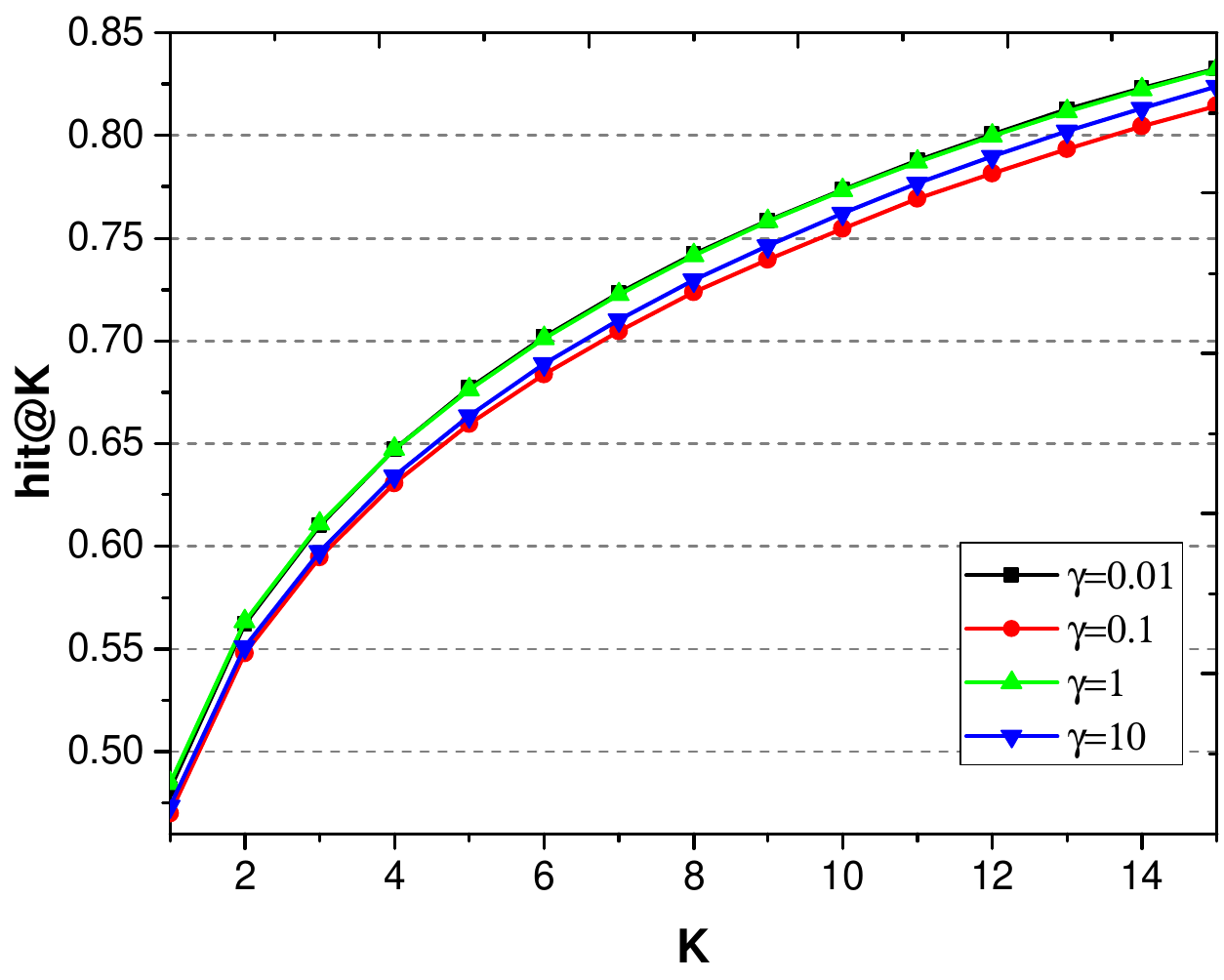}}
\subfigure[ndcg@$K$ on MI]{
\label{fig:gamma-ndcg-mi}\includegraphics[width=0.22\textwidth]{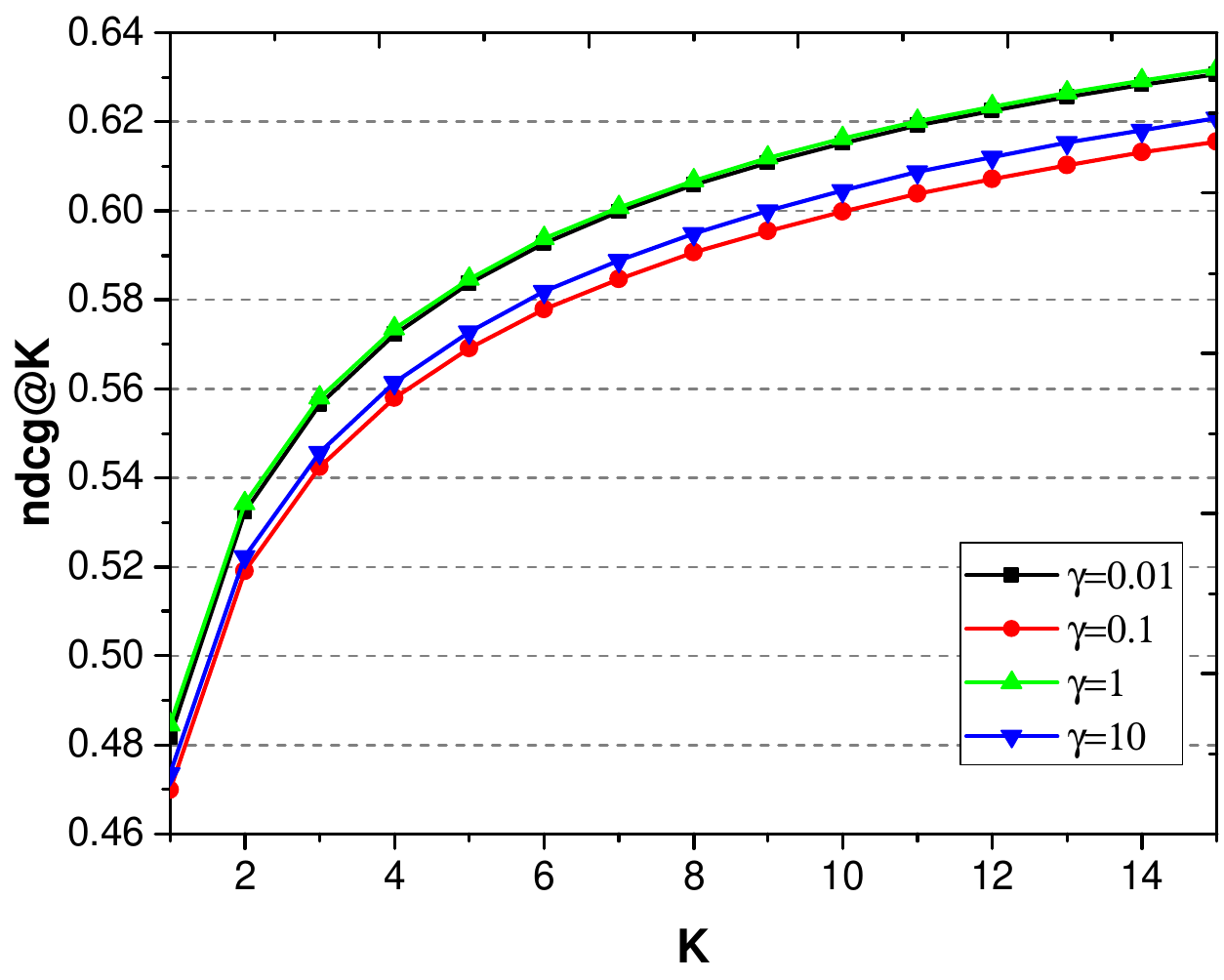}}
\vspace{-15pt}
\caption{Performance comparison of KPRN \wrt $\gamma$ on the MI dataset.}
\label{fig:gamma-effect}\vspace{-15pt}
\end{figure}

Figure~\ref{fig:overall-comparison} reports our experimental results on two datasets \wrt hit@$K$ and ndcg@$K$.
We have the following findings:
\begin{itemize}
\item FMG gives poor performance in both datasets. This indicates that meta-graph based methods, which rely heavily on the predefined meta-graph patterns, may introduce remote entities and fail to fully explore the user-item connectivity.

\item NFM achieves better performance than MF. It makes sense since by treating the rated items as the user features, NFM essentially enhances the second-order user-item proximity ,while MF only considers the first-order user-item connections.

\item Compared to CF-based methods (MF and NFM), the performance of CKE indicates that incorporating KG can solve the data sparsity issue effectively.
In particular, CKE shows consistent improvement over KKBox dataset that is extremely sparse, while only achieving comparable performance to NFM on MI dataset which has denser interaction data.

\item KPRN substantially outperforms CKE \wrt hit@$K$ and ndcg@$K$, achieving the best performance.
By leveraging paths to infer user preference, KPRN is capable of exploring the user-item connectivity in an explicit way, while the embedding-based method (CKE) only utilizes KG to guide the representation learning of items.
This verifies the importance of leveraging both entities and relations of KG.
Further analyzing Figures~\ref{fig:all-ndcg-mi} and~\ref{fig:all-ndcg-kkbox} reveal that KPRN outperforms other baselines by a larger margin \wrt ndcg@$K$, demonstrating the strong capacity of preference ranking.
\end{itemize}

\subsection{Study of KPRN (RQ2)}
To investigate the role of path modeling, we start by explore the influence of relation in paths.
We then study how the weighted pooling operation affects the performance.
\subsubsection{Effects of Relation Modeling}
We consider one variant of KPRN without the relation modeling, termed as KPRN-r.
In particular, the relation embedding $\Mat{r}_{l-1}$ in Equation~\eqref{equ:triple-rep} is discarded to generate the input vector $\Mat{x}_{l-1}$.
In Table~\ref{tab:relation-modeling}, we compare KPRN with KPRN-r in terms of hit@$K$ and ndcg@$K$, where K is selected from $\{5,10,15\}$.
We have the following observations:
\begin{itemize}
\item Without considering relations in paths, the performance of KPRN-r decreases on both datasets. This justifies our intuition that specifying different relations is of importance to capture the path semantics, especially when the same entities are involved.

\item We find that KPRN improves KPRN-r by $6.45\%$ \wrt hit@$5$ on MI, while only $0.70\%$ on KKBox.
One reason may be that as MI is much denser than KKBox and it is common that, in MI, multiple paths connect a user-item pair with similar entities but different relations, whereas fewer paths are offered in KKBox.
This demonstrates that, given strong connectivity between users and items, specifying relations of paths is of more importance to explore the fine-grained interests of users.
\end{itemize}

\subsubsection{Effects of Weighted Pooling}
To integrate the prediction scores of multiple paths between a user-item pair, a weighted pooling operation is carefully designed.
To analyze its effect, we set the value $\gamma$ as $\{0.01,0.1,1,10\}$ and report the performance on MI in Figure~\ref{fig:gamma-effect}.
We find that,
\begin{itemize}
\item When $\gamma$ decrease from $1$ to $0.1$, the weighted pooling operation degrades the performance, since it is similar to max-pooling and selects only the most important paths as the user-item connectivity.

\item The performance \wrt hit@$K$ and ndcg@$K$ becomes poorer, when increasing $\gamma$ from $1$ to $10$. It makes sense since it tends to aggregate contributions from more paths, rather than the most informative ones.
\end{itemize}

\subsection{Case Studies (RQ3)}\label{sec:case-study}

Another desirable property of KPRN is to reason on paths to infer the user preferences towards target items and generate reasonable explanations.
This is because our model capture the higher-level semantics from these key factors: entity, entity type, and relation. 
To demonstrate this, we show an example drawn from KPRN on movie recommendation task.

\begin{figure}[t]
\centering
\includegraphics[width=0.48\textwidth]{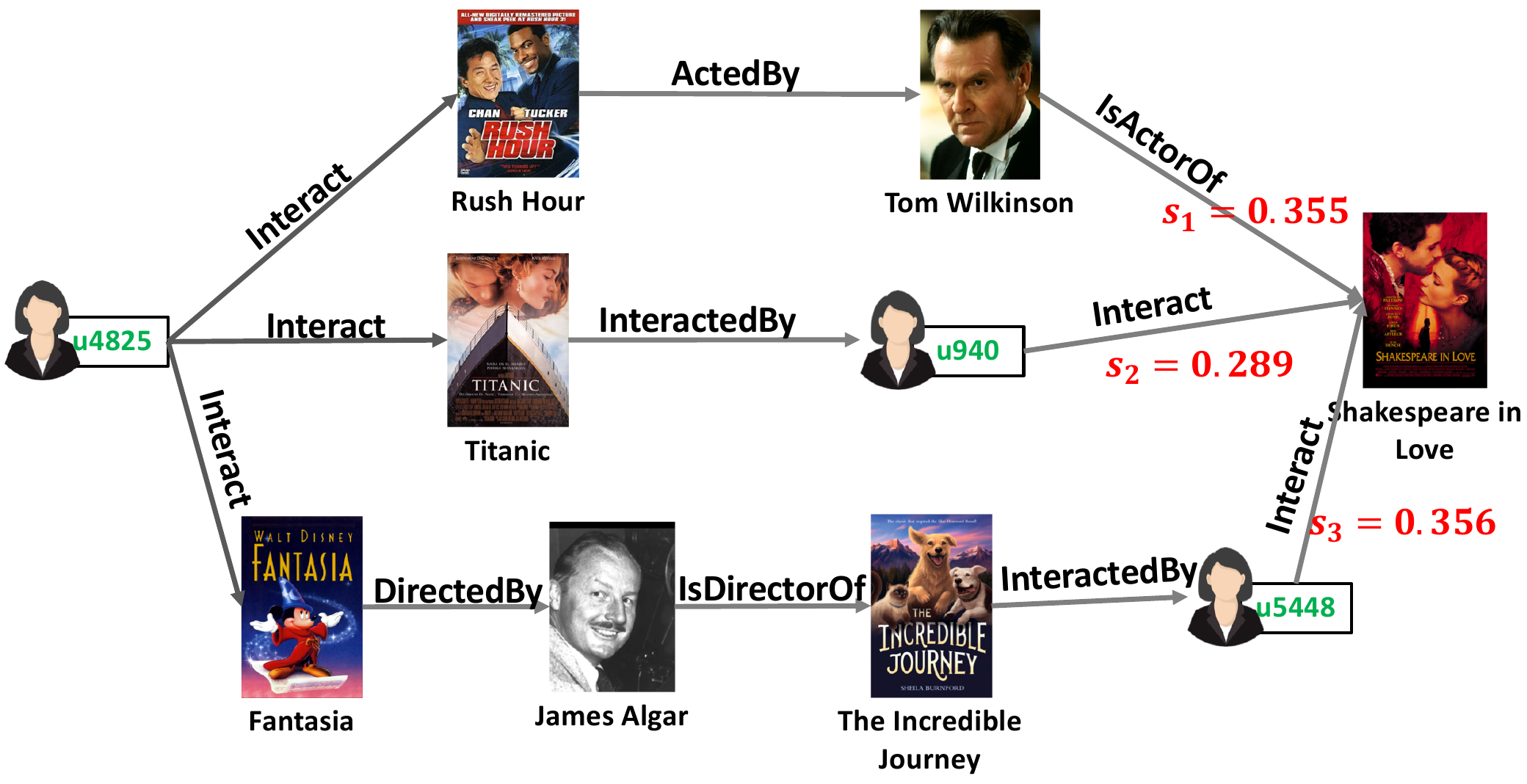}
\vspace{-15pt}
\caption{Visualization of three paths with prediction scores for the user of u4825 in MI dataset. The prediction scores are normalized for illustration.}\label{fig:case-study}
\vspace{-15pt}
\end{figure}

We randomly select a user, whose ID is u4825 in MovieLens-1M, and select the movie Shakespeare in Love from her interaction record.
We then extract all the qualified paths connecting the user-item pair and present the subgraph in Figure~\ref{fig:case-study}.
We have several observations.
\begin{itemize}
\item Collaborative filtering effect plays a pivotal rule to recommend the movie Shakespeare in Love to the user, since the interaction behaviors from other users (\eg u940 and u5448) are involved in two paths.
In particular, the path containing u5448 offers the high contribution score of $0.356$ to infer the user's interest.

\item The target item is connected to what u4825 has watched before (\eg \emph{Rush Hour, Titanic, and Fantasia}) by the shared knowledge entities, such as actor (\emph{Tom Wilkinson}) and director (\emph{James Algar}). This shows that KPRN is capable of extending user interests along KG paths.

\item Analyzing these three paths jointly, we find that different paths describe the user-item connectivity from dissimilar angles, which can be treated as the evidence why the item is suitable for the user.
Specially, we can offer path-wise explanations such as \emph{Shakespeare in Love is recommended since you have watched Rush Hour acted by the same actor Tom Wilkinson} or \emph{since it is similar to Titanic that you watched before.}
This case demonstrates KPRN's capacity of providing informative explanations.
\end{itemize}

\section{Related Work}
Previous solutions on integrating KG into recommendation can be roughly categorized into embedding-based and path-based methods.

\subsection{Embedding-based Methods}
Prior efforts~\cite{DBLP:conf/icml/NickelTK11,CKE,DBLP:conf/recsys/BelliniANS17,DKN} leverage knowledge graph embedding techniques to guide the representation learning of items.
For example, for each item, Zhang~\etal~(\citeyear{CKE}) generated the representation by combining its latent factor from MF and its semantic embedding from TransR.
When performing news recommendation, Wang~\etal~(\citeyear{DKN}) generated the news representation by integrating the knowledge-aware embeddings and word embedding of each word entity within the news content.
More recently, Huang~\etal~(\citeyear{Huang18sigir}) adopted TransE~\cite{TransE} to generate representations for entities and items, and employed memory networks to update the user representations based on her preferences on specific entities.
By exploiting the KG to guide the representation learning, such methods achieve significant improvement in performance.
However, we argue that KGE regularization has not fully explored the connectivity between users and items.
One reason is that the characterization of user-item connectivity is achieved in a rather implicit way.
Moreover, they lack the reasoning ability to infer why a item is recommended for a user.

\subsection{Path-based Methods}
In the literature of path-based methods, some prior studies introduce the connectivity patterns, termed as meta-paths, to explicitly guide the recommender leaning~\cite{DBLP:conf/aaaiss/HeitmannH10,DBLP:journals/sigkdd/SunH12,yu2013collaborative,DBLP:conf/wsdm/YuRSGSKNH14,DBLP:conf/cikm/ShiZLYYW15,DBLP:conf/ijcai/GaoYWZLH18,DBLP:conf/ijcai/GaoYWZLH18}.
Meta-path is defined as a sequence of entity type, such as user-movie-direct-movie, to capture the user-item affinities carried in KG.
For instance, Yu~\etal~(\citeyear{DBLP:conf/wsdm/YuRSGSKNH14}) conducted MF framework over meta-path similarity matrices to perform recommendation.
Such methods, however, use the user-item connectivity to update user-item similarity, but not reason on paths.
Moreover, the performance rely heavily on the quality of meta-paths, which requires domain knowledge.

Several researchers exploit programming models to infer a user preference along paths~\cite{DBLP:conf/recsys/CatherineC16,DBLP:conf/recsys/ChaudhariAM16}.
Nevertheless, these methods fail to learn representations of users and items, thus hardly generalize to unseen interactions.
To solve the issues, recent studies learn the representation for each path~\cite{Meta-path-2018kdd,RKGE,TEM}.
Hu~\etal~(\citeyear{Meta-path-2018kdd}) employed CNN over the embeddings of entities to get a single representation for a path, while recurrent networks are adopted in~\cite{RKGE}.
As such, these methods can combine the strengths of embedding-based and path-based approaches, achieving better performance.
However, the work~\cite{Meta-path-2018kdd} ignores the sequential dependencies of entities and relations within a path; moreover, only entity embeddings are involved in the path modeling~\cite{RKGE}.
Such limitations may hurt the reasoning ability of models.
Towards this end, we propose a model to consider the sequential dependencies, as well as relation semantics, to reason a user-item interaction.

\section{Conclusions}
In this work, we exploit knowledge graph to construct paths as extra user-item connectivity, which is complementary to user-item interactions.
We propose a knowledge-aware path recurrent network to generate representation for each path by composing semantics of entities and relations.
By adopting LSTM on paths, we can capture the sequential dependencies of elements and reason on paths to infer user preference.
Extensive experiments are performed to show the effectiveness and explainability of our model.

In future, we will extend our work in two directions.
First, we attempt to mimic the propagation process of user preferences within KGs via Graph Neural Networks, since extracting qualified paths needs labor-intensive.
Second, as KG links multiple domains (\eg movie and book) together with overlapped entities, we plan to adopt zero-shot learning to solve the cold start issues in the target domain.

\noindent\textbf{Acknowledgement}
This work is supported by NExT, by the National Research Foundation Singapore under its AI Singapore Programme, Linksure Network Holding Pte Ltd and the Asia Big Data Association (Award No.: AISG-100E-2018-002).
Assistance provided by eBay, Search Science Shanghai Director Hua Yang, Manager Xiaoyuan Wu, and intern Mohan Zhang was greatly appreciated.



\bibliographystyle{aaai}
\bibliography{ebay2019}

\end{document}